\documentclass[fleqn,usenatbib]{mnras}

\usepackage{newtxtext,newtxmath}

\usepackage[T1]{fontenc}
\usepackage{ae,aecompl}

\usepackage{graphicx}	
\usepackage{amsmath}	
\usepackage{amssymb}	

\usepackage{hyperref}
\usepackage[switch]{lineno}

\newcommand{\fermi}{\emph{Fermi}}
\newcommand{\source}{S4\,0444+63}
\newcommand{\units}{ph\,cm$^{-2}$\,s$^{-1}$}
\newcommand{\whz}{W\,Hz$^{-1}$}
 
\title[S4~0444+63 radio flare]{A flat-spectrum flare in  S4 0444+63 revealed by a new implementation of multi-wavelength single-dish observations}

\author[M.\ Giroletti \& S.\ Righini]{
M.\ Giroletti,$^{1}$\thanks{E-mail: marcello.giroletti@inaf.it}
and S.\ Righini$^{2}$
\\
$^{1}$INAF Istituto di Radioastronomia, via Gobetti 101, I-40129, Bologna, Italy\\
$^{2}$INAF Istituto di Radioastronomia, Stazione di Medicina, Via Fiorentina 3513, I-40059, Villafontana (BO), Italy
}

\date{Accepted 2019 December 31. Received 2019 December 31; in original form 2019 September 23}

\pubyear{2019}

\begin{document}
\label{firstpage}
\pagerange{\pageref{firstpage}--\pageref{lastpage}}
\maketitle

\begin{abstract}
Relativistic amplification boosts the contribution of the jet base to the total emission in blazars, thus making single dish observations useful and practical to characterise their physical state, particularly during episodes of enhanced multi-wavelength activity. Following the detection of a new gamma-ray source by {\it Fermi}-LAT in July 2017, we observed  S4\,0444+63 in order to secure its identification as a gamma-ray blazar. We conducted observations with the Medicina and Noto radio telescopes at 5, 8, and 24 GHz for a total of 12 epochs between 2017 August 1 and 2018 September 22. We carried out the observations with on-the-fly cross scans and reduced the data with our newly developed Cross-scan Analysis Pipeline, which we present here in detail for the first time.    We found the source to be in an elevated state of emission at radio wavelength,  compared to historical values, which lasted for several months.    The maximum luminosity was reached on 2018 May 16 at 24 GHz, with $L_{24}=(1.7\pm0.3)\times10^{27}\ \mathrm{W\,Hz}^{-1}$; the spectral index was found to evolve from slightly rising to slightly steep.    Besides the new observations, which have proved to be an effective and efficient tool to secure the identification of the source, additional single dish and very-long-baseline interferometry data provide further insight on the physics of the source. We estimate a synchrotron peak frequency $\nu_\mathrm{peak}=10^{12.97}$ Hz and a Doppler factor in excess of $\delta\sim5.0$, with both quantities playing a role in the gamma-ray emission from the source.
\end{abstract}




\section{Introduction}

Blazars are radio loud active galactic nuclei (AGNs) in which the axis of the pair of relativistic jets emerging from the black hole proximity is closely aligned (to within a few degrees) to our observing direction.  This geometry results in a dramatic amplification and shift to high energy of the radiation emitted in the innermost regions, and a shortening of the time scales. As a consequence, the non-thermal jet radiation outshines the contribution of other components in most wavelengths and the overall spectral energy distribution (SED) is dominated by synchrotron radiation at low energy and by inverse Compton emission at high energy, up to gamma rays; hadronic processes may also contribute to this second component.

The physical processes at work in blazars result in electromagnetic radiation that is inevitably variable and broadband. Therefore, their study requires a comprehensive approach considering the different energy and temporal domains. In gamma rays, thanks to its sensitivity and surveying capabilities, the Large Area Telescope (LAT) on board \fermi\ has revealed over 1400 blazars \citep[in the so-called 3LAC sample,][]{3LAC}.  The LAT is continuing to discover new gamma-ray blazars, both thanks to longer integration resulting in a higher significance for weak sources, and to a longer time baseline favouring the detection of new sources undergoing bright flares.  Sources of the former type are of interest as they represent extensions to unexplored domains in low luminosity or high redshift. Those of the latter class are also of great importance since they provide information about the amplitude of the variations in the power of the central engine and the duty cycle of individual objects and of the populations; in turn, this has implications on the detection rates and the blazar contribution to the extragalactic diffuse gamma-ray background.

One outstanding case is the flare observed from \source\ in 2017 \citep{Ciprini2017}. This source was not detected in any of the \fermi-LAT source catalogues up to the third \citep[3FGL,][]{Acero2015}, and was only reported as a rather weak source ($F_{E>100\ \mathrm{MeV}}=(3.8\pm0.6) \times 10^{-9}$ \units) in the fourth \fermi-LAT catalogues of gamma-ray sources \citep[4FGL,][]{4FGL} and of AGNs \citep[4LAC][]{4LAC}.  However, in July 2017 it reached a daily photon flux of $(0.5\pm0.2) \times 10^{-6}$ \units, $\sim130\times$ brighter than in the 4FGL. Had the source reached this value during the first three months of operations of \fermi, it would have been ranked among the top 100 brightest gamma-ray blazars in the sky \citep[the so-called LBAS,][]{Abdo2009}. 

While the high-energy properties of the source will be discussed elsewhere, we here focus on the radio multi-wavelength emission.  We do so thanks to prompt and long-term follow-up observations with the Medicina and Noto 32m radio telescopes. These antennas have been used regularly for continuum blazar observations \citep[e.g.,][]{D'Ammando2013,D'Ammando2014,Raiteri2015,Raiteri2017,Ahnen2017,Larsson2018} but a detailed description of the data acquisition and analysis procedures was not reported yet. This paper therefore serves also as an illustration and a reference of our current technique, which supersedes the previous one based on the on-source/off-source method \citep{Venturi2001}.  We further extend out dataset with 15 GHz observations from the Owens Valley Radio Observatory (OVRO) 40m radio telescope.

The paper is structured as follows: in Sect.\ \ref{sect.observations}, we describe the observations and give details about the new observing and analysis techniques; in Sect.\ \ref{sect.results} we present the results and we discuss them in Sect.\ \ref{sect.discussion}; Sect.\ \ref{sect.conclusions} contains a summary of our conclusions. Appendix \ref{sect.appendix} contains a discussion of the uncertainties and various tests on the accuracy of our procedures.
Radio ($\alpha$) and gamma-ray ($\Gamma$) spectral indexes are defined such that $S_\nu \sim \nu^{-\alpha}$ and $N_E \sim E^{-\Gamma}$, respectively.

\section{Observations and data reduction}
\label{sect.observations}

\begin{table}
	\centering
	\caption{Telescope configurations and features}         
	\label{t.telescopes}    
	\begin{tabular}{l c c c c}        
		\hline
		Telescope & Frequency   & Beamsize & Tsys$^{(a)}$  & Max Gain \\ 
		          & band (GHz)  & (arcmin) & (K)   & (K/Jy)   \\ 
		\hline                       
		Noto      & 4.65 - 5.02 & 7.5      & 30    & 0.16     \\
		Medicina  & 8.18 - 8.86 & 4.9      & 38    & 0.14     \\
		Medicina  & 23.5 - 24.7 & 1.6      & 60    & 0.11     \\
		\hline
	\end{tabular}\\
$^{(a)}$ Tsys pointing at Zenith, with $\tau_0=0.1$
\end{table}

Observations were carried out with the 32-m dishes located in Medicina and Noto; both instruments are owned and managed by INAF (National Institute for Astrophysics, Italy). 
Continuum acquisitions were performed exploiting On-The-Fly (OTF) cross-scans in Equatorial coordinates. Table \ref{t.telescopes} lists the telescope main features and employed configurations, while Table \ref{t.scans} provides the scanning parameters used during the observations. The on-source integration time associated to each flux density measurement is 37.5\,s at 5~GHz, 40.0\,s at 8~GHz and 52.5\,s for 24~GHz acquisitions.   
Flux density calibration was carried out observing 3C\,123, whose reference flux density was computed, for the observed band central frequency, according to \citet{Perley2013}. 
For 24-GHz observations, the atmospheric contribution was also taken into account in the calibration procedure; zenithal opacity was estimated by means of skydip acquisitions. 

\begin{table}
\centering                          
\caption{Cross-scan parameters}         
\label{t.scans}    
\begin{tabular}{l c c c}        
\hline                 
Frequency   & Scan length & Scan speed  & Sampling \\ 
(GHz)  & (degrees)    & (arcmin/s)  & (s)      \\ 
\hline                       
5 & 1.0      & 4.0    & 0.040     \\
8 & 0.6      & 2.4    & 0.040     \\
24& 0.2      & 0.8    & 0.040     \\
\hline                                  
\end{tabular}
\end{table}

All the recordings, and the following analysis operations, are structured according to the scan-subscan definition implemented in the antenna control system. For cross-scans, we define subscan the single segment acquired moving the antenna at constant speed in a single direction along one of the two orthogonal axes. A scan is, instead, a group of a multiple of four subscans, as the basic unit allowed by the system is composed by two latitude (Dec) and two longitude (RA) subscans. The acquisition system produces one folder for each scan, containing one FITS file for each subscan.     

\begin{figure*}
\includegraphics[width=0.67\textwidth]{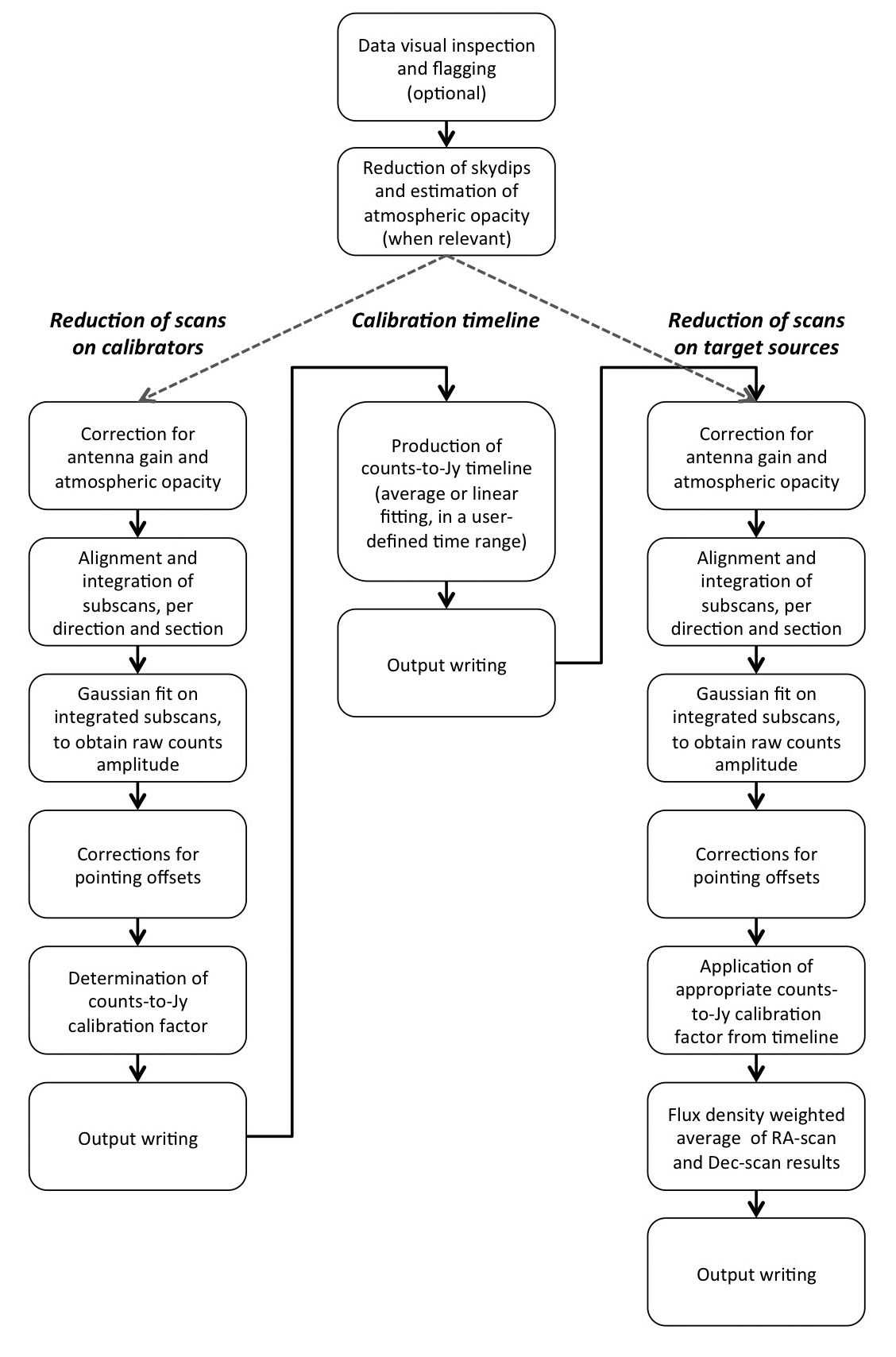}
\caption{Conceptual scheme of the Cross-scan Analysis Pipeline.}
\label{fig.capscheme}%
\end{figure*}

Data reduction was performed using CAP (Cross-scan Analysis Pipeline, \url{https://github.com/discos/CAP}), a set of routines we wrote in IDL$\copyright$ (programming language by Exelis). 
The pipeline consists in a sequence of operations, structured as follows and illustrated in Fig.~\ref{fig.capscheme}. 

\begin{description}  
\item [\textit{Data rearranging}] The dataset is rearranged so as acquisitions are separated by frequency and by the different purpose of the observed sources (calibrators, skydips, targets); 
\item [\textit{Data flagging}] Optionally, users can launch a GUI to visually inspect each subscan and interactively assign flags to the data, in order to assess which acquisitions are to be taken into account in the subsequent analysis phase. As the FITS files contain both the Left Circular Polarisation (LCP) and Right Circular Polarisation (RCP) data streams - which might greatly differ from one another, due to the asymmetric impact of interferences (RFI) - users can choose whether to accept or reject the single polarisations. If this step is skipped, the following actions are carried out using the whole dataset; 
\item [\textit{Skydips reduction}] When relevant, i.e. at frequencies above 10 GHz, skydip acquisitions are performed in order to estimate the atmospheric opacity. At this stage, such data are fitted according to proper atmospheric models, exploiting some telescope-related parameters (e.g. the receiver temperature) and the weather parameters recorded during the acquisitions. The procedure returns a table listing the MJD-tagged estimates for $\tau_0$ (zenithal opacity) obtained from all the available skydips; 

\begin{table*}
\centering                          
\caption{Results of observations}         
\label{t.results}    
\begin{tabular}{l l c c c c c c }        
\hline            
\multicolumn{2}{c}{Date}   &  $S_5$ & $S_8$  & $S_{24}$ & $\alpha_{5-8}$ & $\alpha_{8-24}$ \\
(yyyy-mmm-dd) & (MJD)  & (Jy) & (Jy) & (Jy)  \\ 
\hline                       
2017 Aug 1 & 57966.4 & $0.69\pm0.01$ & $0.79\pm0.01$ & $0.98\pm0.04$ & $-0.26\pm0.04$ & $-0.21\pm0.04$ \\
2017 Oct 14 & 58040.1 & \dots         & $0.83\pm0.02$ & $0.94\pm0.03$ & $0.15\pm0.33^a$ & $-0.12\pm0.05$ \\
2017 Oct 15 & 58041.9 & $0.90\pm0.16$ & \dots         & \dots & \dots & \dots \\
2018 Jan 1 & 58136.8 & \dots         & $0.93\pm0.03$ & $0.97\pm0.04$ & \dots & $-0.04\pm0.05$ \\
2018 Feb 10 & 58159.8 & \dots         & $0.88\pm0.03$ & $0.95\pm0.07$ & \dots & $-0.07\pm0.08$ \\
2018 Mar 10 & 58187.6 & \dots         & $0.86\pm0.02$ & \dots & \dots & \dots \\
2018 Apr 6 & 58214.6 & $0.81\pm0.04$ & \dots         & \dots  & \dots & \dots \\
2018 Apr 16 & 58224.5 & \dots         & $0.85\pm0.06$ & $0.91\pm0.05$ & \dots & $-0.07\pm0.09$   \\
2018 May 16 & 58254.7 & \dots         & $0.84\pm0.04$ & $1.02\pm0.20$ & \dots & $-0.18\pm0.20$   \\
2018 Jun 21 & 58290.5 & \dots         & $0.82\pm0.02$ & $0.82\pm0.06$ & \dots & $0.01\pm0.08$   \\
2018 Jul 25 & 58324.4 & $0.71\pm0.07$ & \dots         & \dots  & \dots & \dots \\
2018 Sep 22 & 58383.2 & \dots         & $0.74\pm0.02$ & $0.63\pm0.05$ & \dots & $0.15\pm0.09$   \\
\hline     
\end{tabular}\\
$^{(a)}$ $\alpha_{5-8}$ calculated using data in two consecutive days (58040-58041)
\end{table*}
\item [\textit{Calibrators reduction}] A scan-based integration of the acquisitions on flux density calibration sources is performed. The green-lighted subscans, i.e. the ones positively flagged, are properly aligned and integrated. Prior to being averaged, they are corrected for the antenna gain curve and, when due, for the atmospheric opacity. Both parameters are Elevation-dependent, thus each subscan is individually handled. A Gaussian fitting is finally performed on the integrated results, so as to measure the signal amplitude in raw counts. Users can decide whether they want the fitting to assume a linear or cubic baseline, or have both the options carried out in parallel. The measured amplitude is then corrected for the pointing error - which is measured thanks to the availability of the orthogonal RA-Dec subscans - as the primary beam response as a function of the pointing error is analytically known a priori. At this point, the source theoretical flux density is computed, taking into account the actual observed band, according to  \citet{Perley2013}, and such flux density is divided by the raw amplitude, in order to obtain a counts-to-Jy conversion factor;     
\item [\textit{Conversion timeline}] Once all the data acquired on flux density calibrators are processed, it is important to decide how to interpolate the counts-to-Jy conversion factors, as they might have been measured during long observing sessions. Ideally, as the main antenna-related and weather-dependant effects were compensated in the previous phase, the conversion factors should be constant. However, fluctuations might be present if instabilities affected the weather or the instrument. For this reason, users can decide to average or linearly interpolate the conversion factors, choosing a time window within which to perform the desired operation;   
\item [\textit{Targets reduction}] From a conceptual point of view, this procedure performs the same operations inserted in the "calibrators reduction" phase. Only, this time the raw amplitude produced by the Gaussian fitting is multiplied by the proper calibration factor - chosen from the counts-to-Jy timeline previously listed. The final product is a table with the measured flux densities, one for each original scan.  

\end{description}

\begin{figure}
    \includegraphics[width=\columnwidth]{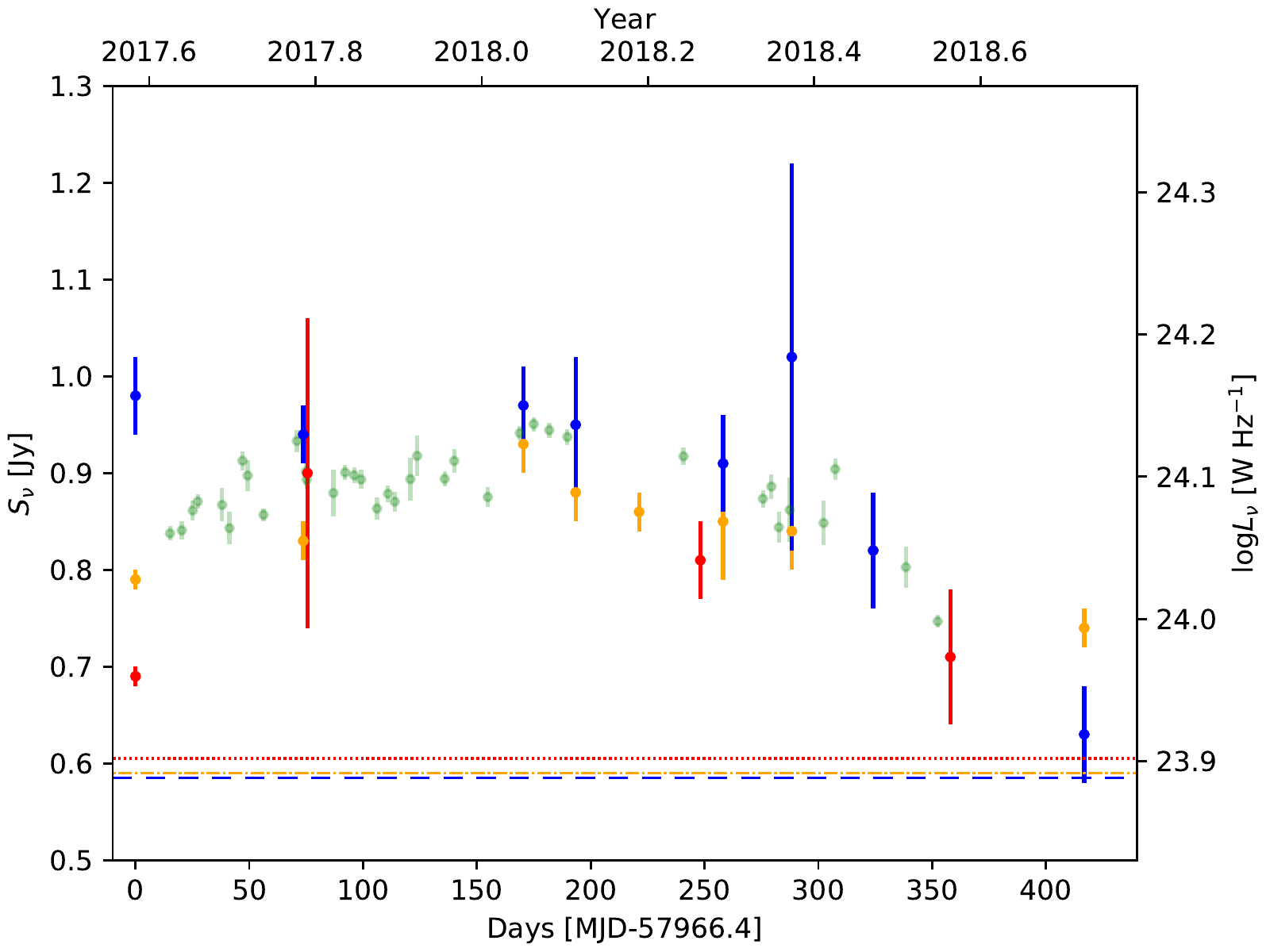}
\caption{Light curve of \source\ during our campaign at 5 (red points), 8 (orange), and 24 (blue) GHz; OVRO 15 GHz data are also shown in background as pale green dots. The horizontal lines report historical flux density taken from NED at the corresponding frequency: red dotted line: 5 GHz \citep{Becker1991}; dot-dash orange line: 8 GHz \citep{Rickett2006}; blue dashed line: 24 GHz, as interpolated between 15 \citep{Richards2011} and 30 GHz \citep{Lowe2007}.}
              \label{fig.lightcurve1}%
\end{figure}

All along the process, LCP and RCP data streams are kept separate, so they produce independent flux density measurements. For a given polarisation, though, the two measurements deriving from the RA and Dec integrated semi-scans are processed via a weighted average, and only one flux density is finally obtained. 
Partial results are listed in a separate table, and users can optionally inspect the results achieved by reducing each of the original subscans. 

The treatment of uncertainties and examples of the tests carried out to assess the software performance are given in Appendix \ref{sect.appendix}.

\begin{table}
\centering                          \small
\caption{Average values}         
\label{t.averages}    
\begin{tabular}{l l}        
\hline
Quantity & Value \\
\hline                       
$\langle S_5 \rangle \pm \sigma_{S_5}$ & $(0.70\pm0.10)$ Jy\\
$\langle S_8 \rangle \pm \sigma_{S_8}$ & $(0.81\pm0.05)$ Jy  \\
$\langle S_{15} \rangle \pm \sigma_{S_{15}}$ & $(0.88\pm0.04)$ Jy  \\
$\langle S_{24} \rangle \pm \sigma_{S_{24}}$ & $(0.91\pm0.12)$ Jy  \\
\hline
$\langle L_5 \rangle \pm \sigma_{L_5}$ & $(1.18\pm0.16) \times10^{27}\, \mathrm{W\,Hz}^{-1}$ \\
$\langle L_8 \rangle \pm \sigma_{L_8}$ & $(1.38\pm0.09) \times10^{27}\, \mathrm{W\,Hz}^{-1}$ \\
$\langle L_{15} \rangle \pm \sigma_{L_{15}}$ & $(1.49\pm0.07) \times10^{27}\, \mathrm{W\,Hz}^{-1}$ \\
$\langle L_{24} \rangle \pm \sigma_{L_{24}}$ & $(1.53\pm0.21) \times10^{27}\, \mathrm{W\,Hz}^{-1}$ \\
\hline
$\alpha_{5-8}^{}$ & $-0.29\pm0.30$ \\
$\alpha_{8-15}^{}$ & $-0.14\pm0.14$ \\
$\alpha_{15-24}^{}$ & $-0.06\pm0.31$ \\
\hline
$V_5$ & 0.028 \\
$V_8$ & 0.084 \\
$V_{15}$ & 0.112 \\
$V_{24}$ & 0.093 \\
\hline     
\end{tabular}
\end{table}


\section{Results}
\label{sect.results}

\subsection{New observations}
\label{sect.newobservations}

In Table \ref{t.results}, we give the flux density and the spectral index observed at each epoch.  Since the observations in Noto (at 5 GHz) typically occurred on different dates than those in Medicina, we only have two near-simultaneous $\alpha_{5-8}$ values, while in general for each Medicina observation we also have the associated $\alpha_{8-24}$.  The flux density trend is also shown graphically in Figure \ref{fig.lightcurve1}, where we also report concurrent 15 GHz data points from the OVRO 40m radio telescope (see Sect.~\ref{sect.archivaldata}) and reference historical values. Flux densities are converted to luminosities (reported on the right hand side $y$-axis) based on the redshift $z=0.781$ reported by \citet{Stickel1993} and on the latest cosmological parameters \citep{Planck2018}.  For each frequency, we also give the weighted mean and the standard deviation of the measurements in Table~\ref{t.averages}; from the mean values, we also calculate the average spectral index.   

Our observations cover a period of over a year, between 2017 July 13 and 2018 September 22 (MJD $ 57966-58383 $). Considering the period of our observations, flux densities range between 0.63 and 1.02 Jy, corresponding to a monochromatic luminosity in the range $1.1\times10^{27} \le L_r/(\mathrm{W\,Hz}^{-1}) \le 1.7\times10^{27}$.  These values are consistently above the historic state of \source\ at the same frequencies, as well as, e.g., the NVSS luminosity $L_{1.4}=7.2 \times 10^{26}\, \mathrm{W\,Hz}^{-1}$  ($L_r=1.0\times10^{43}\, \mathrm{erg\,s}^{-1}$).


Besides the enhanced activity respect to historical values, our observations allow us to reveal the variability on monthly time scales following the gamma-ray flare (see Fig.~\ref{fig.lightcurve1}), even accounting for the significant observational uncertainties associated with each measurement.  In particular, there are indications of an overall trend of decreasing activity as a function of time at 24 GHz. At lower frequency, the smaller error bars allow us to observe an initial rise followed by a decay in the 8 GHz data.   The 5 GHz observations are consistent with such trend, albeit in a less compelling way because of the more limited number of epochs. Quantitatively, we can characterise the variability through the variability index $V$, defined as:
\begin{equation}
    V=\frac{(S_\mathrm{max}-\sigma_\mathrm{max})-(S_\mathrm{min}+\sigma_\mathrm{min})}{(S_\mathrm{max}-\sigma_\mathrm{max})+(S_\mathrm{min}+\sigma_\mathrm{min})}
\end{equation}
For our observations, $V$ grows as a function of frequency, from $V_5=0.028$ to $V_{24}=0.093$ (see also Table~\ref{t.averages}, where the slightly larger value found at 15 GHz is due to the smaller uncertainties for the OVRO data).

Formally, the peak luminosity was reached on MJD 58254 at 24 GHz, with $L_{24}=(1.7\pm0.3)\times10^{27}\ \mathrm{W\,Hz}^{-1}$. However, the data taken on MJD 58254 are those most affected by the weather, resulting in the largest uncertainty, so we point out the possibility that the observed maximum was reached on MJD 58136.  This provides a radio-gamma delay  $\Delta t=191$ days, assuming the onset of the gamma-ray activity to have taken place on 2017 July 11 (MJD 57945). We also performed a second-degree polynomial fit to the densely sampled 15 GHz light curve, which provides a maximum at $\Delta t=165\pm35$ days, in agreement with what estimated through the 24 GHz observations.

The spectral index is overall flat-to-inverted, with values of $\alpha_{5-8}=-0.29\pm0.30$ and $\alpha_{15-24}=-0.06\pm0.31$ when calculated using the average flux densities during the observations (see Table~\ref{t.averages}). While consistent with the reference state of the source calculated using archival data, these values tend to indicate a more inverted spectrum then usual.  In agreement with the time evolution discussed in the previous paragraph, we can also point out a mild evolution of the high frequency spectral index; in particular, it starts from inverted ($\alpha_{8-24}=-0.21$ during the first observations, and generally $\alpha < 0$ until MJD 58254) to flat/positive at the end of the campaign.

\begin{figure}
    \includegraphics[width=\columnwidth]{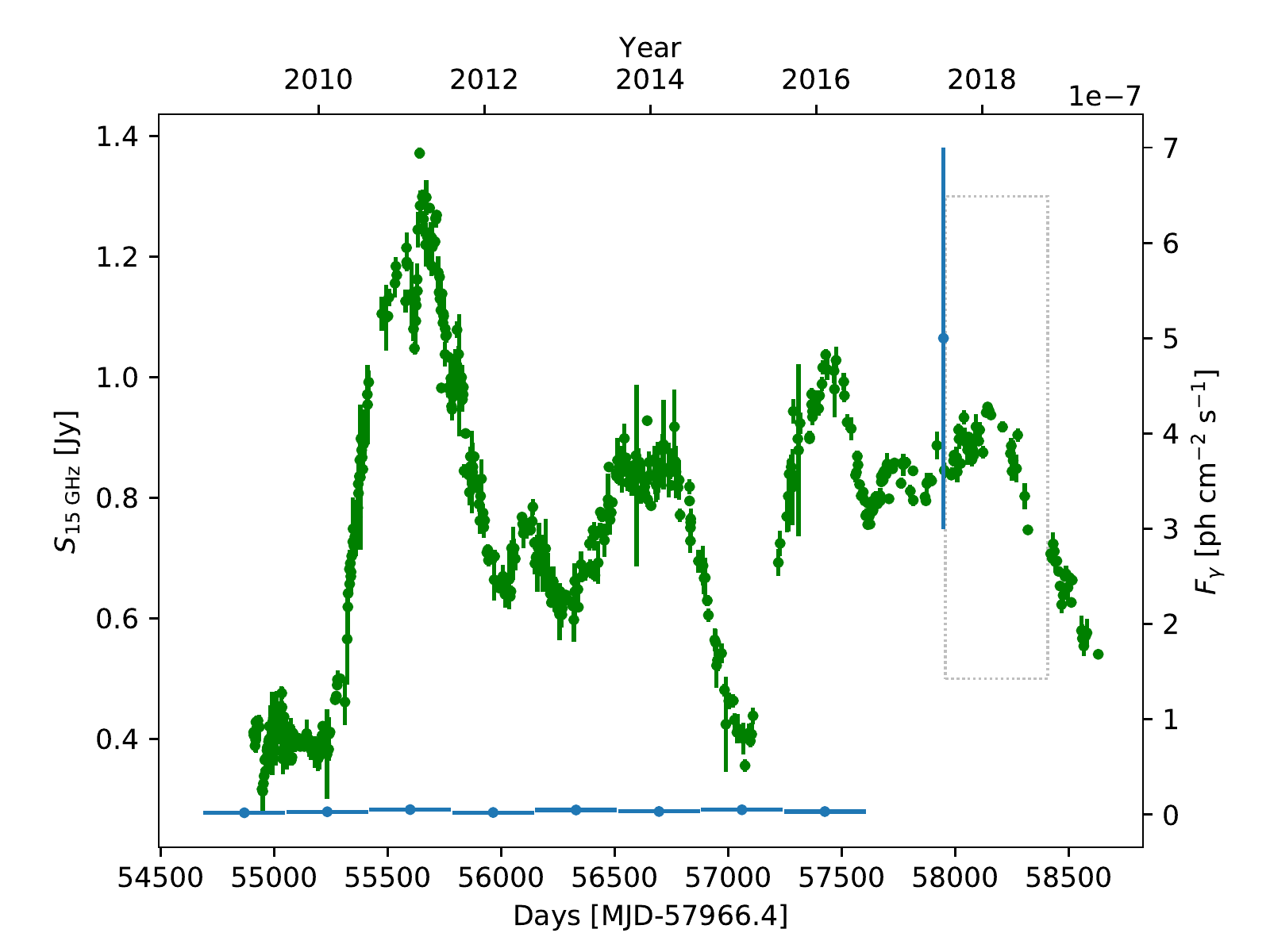}
\caption{Long term light curve of \source\ from OVRO (green dots) overlaid to the \fermi\ gamma-ray data (blue points) in yearly bins (from the 4FGL) and during the July 2017 flare \citep{Ciprini2017}; the grey box shows the area illustrated in Fig.~\ref{fig.lightcurve1}}
              \label{fig.lightcurve2}%
\end{figure}

Our observations thus reveal that \source\ has been undergoing an episode of enhanced activity at radio wavelengths starting simultaneously to its July 2017 gamma-ray flare and evolving over about a year.  This finding lends support to the initial suggestion that \source\ could be a new gamma-ray and X-ray blazar \citep{Ciprini2017}.   

\subsection{Archival data}
\label{sect.archivaldata}

As part of a regular monitoring of more than 1800 blazars, the Owens Valley Radio Observatory (OVRO) 40m radio telescope has been observing \source\ at 15 GHz about twice per week since 2008 \citep{Richards2011}. The OVRO data are monochromatic but the duration of the project makes them very worthwhile to provide a context for the behaviour of the source during our campaign, as illustrated by the light curve shown in Figure \ref{fig.lightcurve2}. In the same plot, we also show yearly values for the \fermi\ data between August 2008 and 2016 \citep{4FGL}, plus the elevated state of gamma-ray activity observed in July 2017 \citep{Ciprini2017}.

The long term light curve reveals that the enhanced activity started in 2017 is the last of a longer series of five events.  Most of these events show much higher flux density increments, in particular the one culminating at the beginning of 2011 corresponds to a rise of the flux density from $S_{15}\sim0.4$~Jy to $S_{15}\sim1.3$~Jy over about one year.  Remarkably, the \fermi-LAT data do not shown any hint of high energy variability during these episodes of radio activity.  Overall, the 15 GHz variability index grows to $V_{15}=0.59$ when the entire OVRO data set is considered.

\section{Discussion}
\label{sect.discussion}
   
\subsection{On the association of the radio and gamma-ray sources}\label{sect.identification}

The accuracy of the localisation of gamma-ray sources by \fermi-LAT is of order of a few arc minutes, with a dependence on flux and photon index.  Therefore, the number of possible low-frequency counterparts positionally consistent with any gamma-ray source is generally very large.  The \fermi-LAT collaboration usually adopts two methods to claim high-statistical-significance \textit{associations} between the gamma-ray sources and objects, one applying Bayes' Theorem to catalogues of known classes of gamma-ray emitters, the other based on the $\log S-\log N$ relation for surveys of radio and X-ray sources. However, a secure \textit{identification} is only claimed on the basis of either spatial coincidence of extended sources \citep[e.g., in the case of the radio galaxy Centaurus A,][]{Abdo2010} or a timing coincidence, e.g.\ through the measurement of the rotation period in pulsars.  For blazars, \textit{identification} is established when correlated variability in gamma rays and at lower frequency is reported.  In the 3FGL, this has only been possible for the 0.8\% of the sources (with only 26 blazar identification out of the 3033 3FGL sources, despite blazars being the most common class of gamma-ray sources).
Lacking significant gamma-ray variability in the first eight years of the \fermi\ mission, \source\ was not detected up to the 3FGL and only associated on a statistical basis to the 4FGL source J0442.7+6142, with a probability of 99.4\% using the Bayesian method and 85.3\% using the likelihood method \citep{4LAC}.

The bright gamma-ray flaring in July 2017 and the following multi-wavelength radio monitoring campaign provide an opportunity to investigate correlated variability.  The inspection of the long term OVRO light curve shows that the radio enhancement started after the gamma-ray flare in July 2017 is not outstanding with respect to other episodes of variability.  The added value of multi-frequency information obtained by the INAF radio telescopes however indicates that this episode of activity is characterised by a radio spectrum $\alpha=-0.26$ immediately after the gamma-ray detection. This value is more inverted than any available historical or later measurement, a behaviour that is a classical signature of efficient particle acceleration and a compact emission region.  Since none of the radio flares seen by OVRO has a comparable multi-wavelength coverage, it is not possible to quantify how exceptional this flare is.  However, the coincidence of the 2017 gamma-ray flare and the subsequent radio spectral evolution provide a very consistent picture in which \source\ is responsible for the gamma-ray emission.

\subsection{Physics of \source}

Having established the classification of \source\ as a gamma-ray blazar, we consider this as an ideal case to investigate some physical characteristics that were otherwise derived only on a statistical basis for large samples of gamma-ray blazar associations.  For instance, \citet{Liodakis2018} studied the distribution of the delay between gamma-ray flares and the radio response in a sample of bright blazars, finding median time lags of $\sim 100-160$ days.  Our measurement of  $\Delta t\sim160-190$ days is in the upper range of what reported by \citet{Liodakis2018}, placing the radio emitting region substantially downstream of the gamma-ray one.

   \begin{figure}
   \centering
   \includegraphics[width=\hsize]{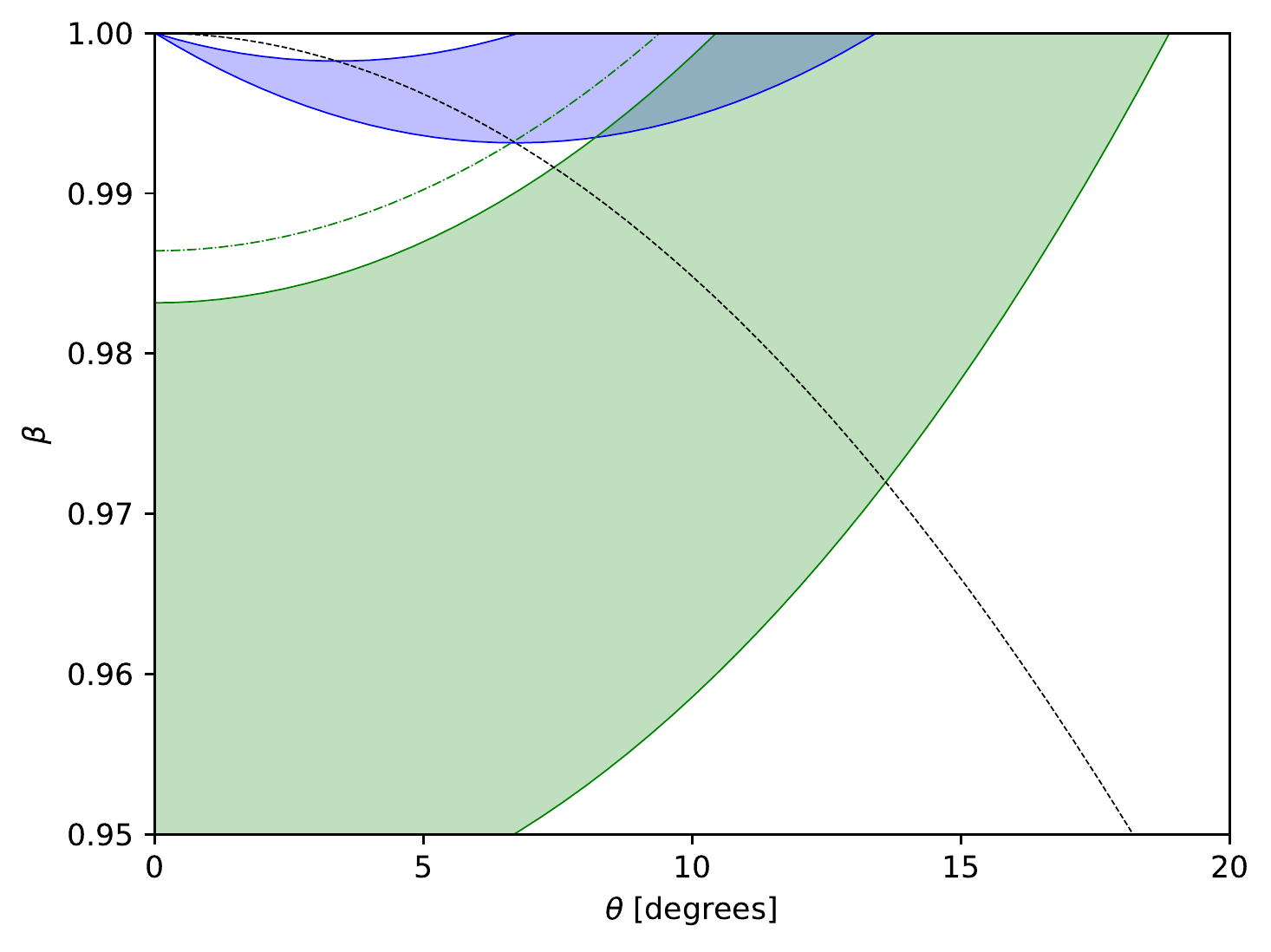}
      \caption{$(\theta,\beta)$ plane for \source.  The blue and green shaded areas indicate the regions allowed by the proper motion and the core dominance arguments, respectively.  The thin  dashed black line corresponds to $\Gamma=1/\sin \theta$; the thin dot-dash green line shows a solution for a total radio luminosity reduced by a factor $3\times$ (see Sect.~\ref{sect.dutycycle}). 
              }
         \label{fig.theta_beta}
   \end{figure}

Moreover, LAT-detected blazars have been reported to have on average higher Doppler factors than non-LAT-detected blazars, with statistically significant differences in the viewing angle distributions between gamma-ray bright and weak sources \citep{Savolainen2010}.  We can discuss the kinematics of \source\ by considering the work of \citet{Britzen2008}: a large project studying bright compact radio sources with Very Long Baseline Interferometry (VLBI).  \citet{Britzen2008} considered three epochs between 1992/07/27 and 1996/08/19, identifying a total of five components: a core and four jet knots (within 6 mas from the core), three of which detected at all three epochs.  The innermost feature showed a proper motion with apparent superluminal velocity $\beta=v/c=(12.7\pm1.4)$.  This provides a first direct lower limit on the jet Lorentz factor $\Gamma \ge 11.3$.  The inspection of the published VLBI images of \source\ \citep{Taylor1994,Britzen2008} allows us to also determine a lower limit on the jet/counter-jet brightness ratio $R_\mathrm{J-CJ}\ge205$, which provides and additional (looser) constraint on the Lorentz factor $\Gamma\ge 1.63$ and viewing angle $\theta\le38^\circ$.

The source is in the LOw-frequency Radio CATalog (LORCAT) of flat-spectrum sources assembled by \citet{Massaro2014}. LORCAT contains sources detected at both 325 MHz in the WENSS and at 1.4 GHz in the NVSS, and with a threshold  $\alpha_\mathrm{0.3-1.4}\le 0.4$; \source\ has a WENSS-NVSS spectral index $\alpha_\mathrm{0.3-1.4} = 0.23 \pm 0.02$.  We interpolate these data to determine the source total flux density and power at 408 MHz, which are $S_{0.4}=570$ mJy and $P_\mathrm{tot}=1.1\times10^{27}$ \whz, respectively.  In turn, we can use the latter value to calculate the expected intrinsic value of the radio core power, based on the well-know correlation studied by \citet{Giovannini1988,Giovannini2001}. By comparing this value with the observed one \citep[$P_\mathrm{c,\ obs}=5.4\times10^{26}\ \mathrm{W\,Hz}^{-1}$, based on $S_\mathrm{5,\ VLA}=321$ mJy reported by][]{Taylor1996}, and allowing for a factor $3\times$ variability, we can derive further constraints on the amount of relativistic boosting of the radio core emission.

These constraints are shown graphically in Fig.~\ref{fig.theta_beta}, along with the $\Gamma=1/\sin\theta$ line.  The overlap between the green and blue shaded regions indicates the region of $(\theta,\beta)$ values consistent with both arguments (the jet-counterjet brightness ratio yields only looser constraints that are not shown in the plot). The vertex nearest to the $\Gamma=1/\sin\theta$ relation has coordinates $(\theta,\beta)=(9.4^\circ,0.9965)$, corresponding to a Doppler factor $\delta\sim5.0$.

   \begin{figure}
   \centering
   \includegraphics[width=\hsize]{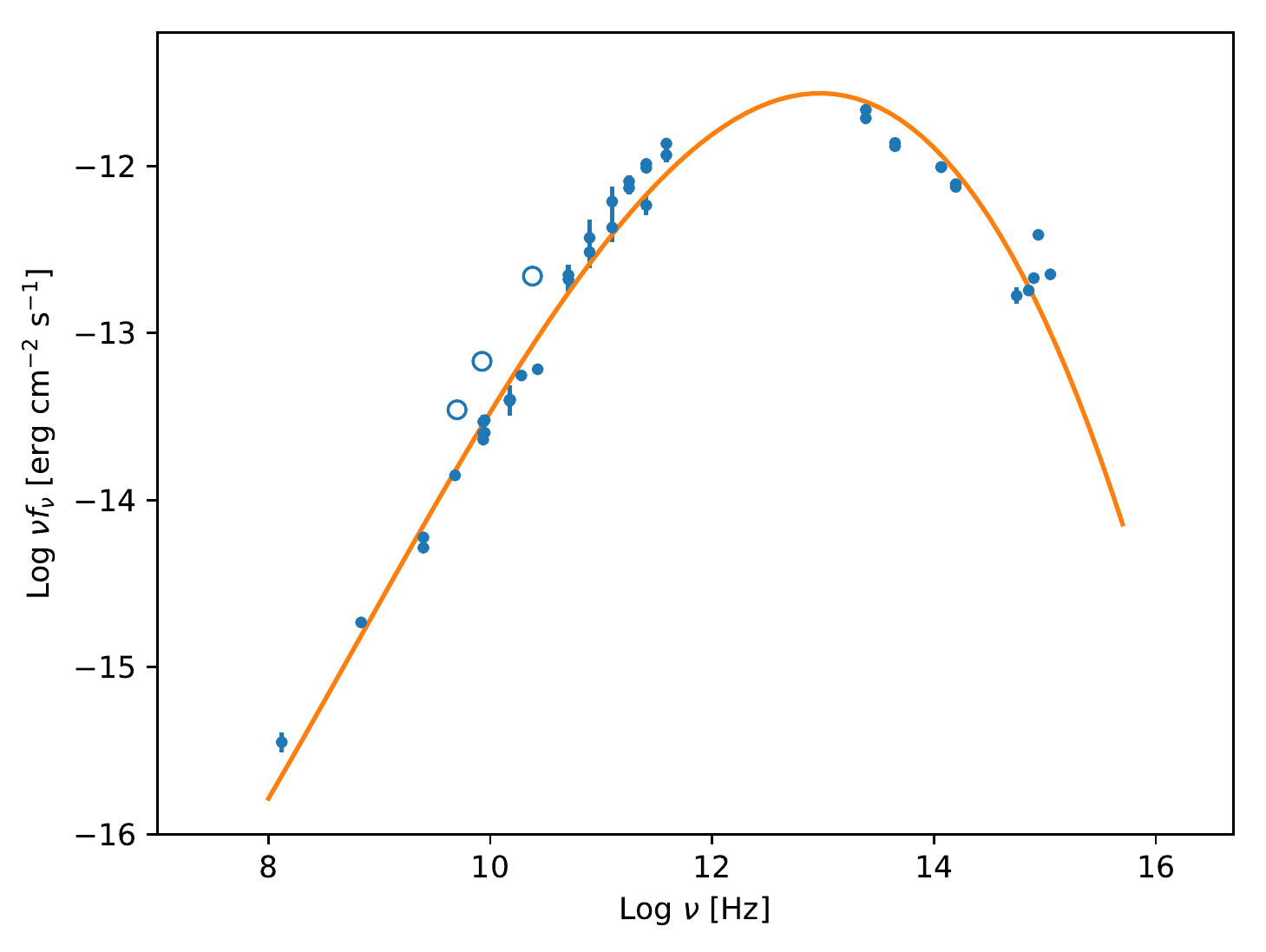}
      \caption{Rest frame SED of \source\ based on data from NED (blue dots), overlaid to a polynomial fit $\log(\nu f_\nu) = a + b\log\nu + c(\log\nu)^2 + d(\log\nu)^3$, with $a=-9.7$, $b=-4.1$, $c=0.60$, $d=-0.023$, calculated with the ASI-SSDC SED builder v.~3.2 (\url{https://tools.ssdc.asi.it/SED/}).  Data from the new observations are also shown as empty circles but were not considered in the fit.
              }
         \label{fig.sed}
   \end{figure}

   
\subsection{Activity duty cycle of FSRQs}\label{sect.dutycycle}
\source\ was discovered within the fourth zone \citep[S4,][]{Pauliny1978} of the Strong Source surveys. Differently from e.g. the 3C and B2 surveys, which were done in the 100's MHz domain, the S4 transit scans were carried out at 5 GHz, making it more suitable to discover flat spectrum sources. The survey includes 269 sources at $S > 0.5$ Jy and it is essentially complete above such threshold; the sources were then followed up with the Effelsberg 100m radio telescope at 2.7, 5.0, and 10.7 GHz.  Historically \citep[see e.g.][]{Blandford2019}, the S4 has therefore been a turning point highlighting the importance of flat-spectrum sources, in particular as a key ingredient for the unification of radio galaxies and quasars. 

Bright, flat-spectrum radio emission is the signature of Doppler-boosted relativistic jet emission and it is considered a defining feature for the search of low-frequency counterparts to gamma-ray sources. Several works have shown the existence of a highly significant correlation between radio and gamma-ray emission in blazars \citep{Ghirlanda2010,Mahony2010,Ackermann2011}.  However, it is to be noted that while radio flux-density provides an indication of the maximum level of gamma-ray emission possible for a source, also lower values are allowed, including the possibility that no gamma rays at all are observed even from very radio-bright blazars \citep{Ackermann2011}. Initially, \citet{Ghirlanda2011} showed that only a minor fraction ($\sim1/15$) of radio sources of the AT20G survey were detected in gamma rays by \fermi-LAT after 11 months.  Even considering longer integrations and higher radio flux densities, \citet{Lister2015} still found that 23\% of the brightest blazars in the northern sky were not detected by the \fermi-LAT in the 3FGL period. \citet{Ghirlanda2011} concluded that a significant decadal variability in gamma rays was necessary to account for the non-detection of the radio brightest blazars (with long-term gamma-ray flux variations described by a lognormal probability distribution with standard deviation $\sigma \ge 0.5$). \citet{Lister2015} considered also the SED and the beaming properties of the sources in their sample and ascribed the non-detections to a combination of instrumental effects and low Doppler factors. 

   \begin{figure}
   \centering
   \includegraphics[width=\hsize]{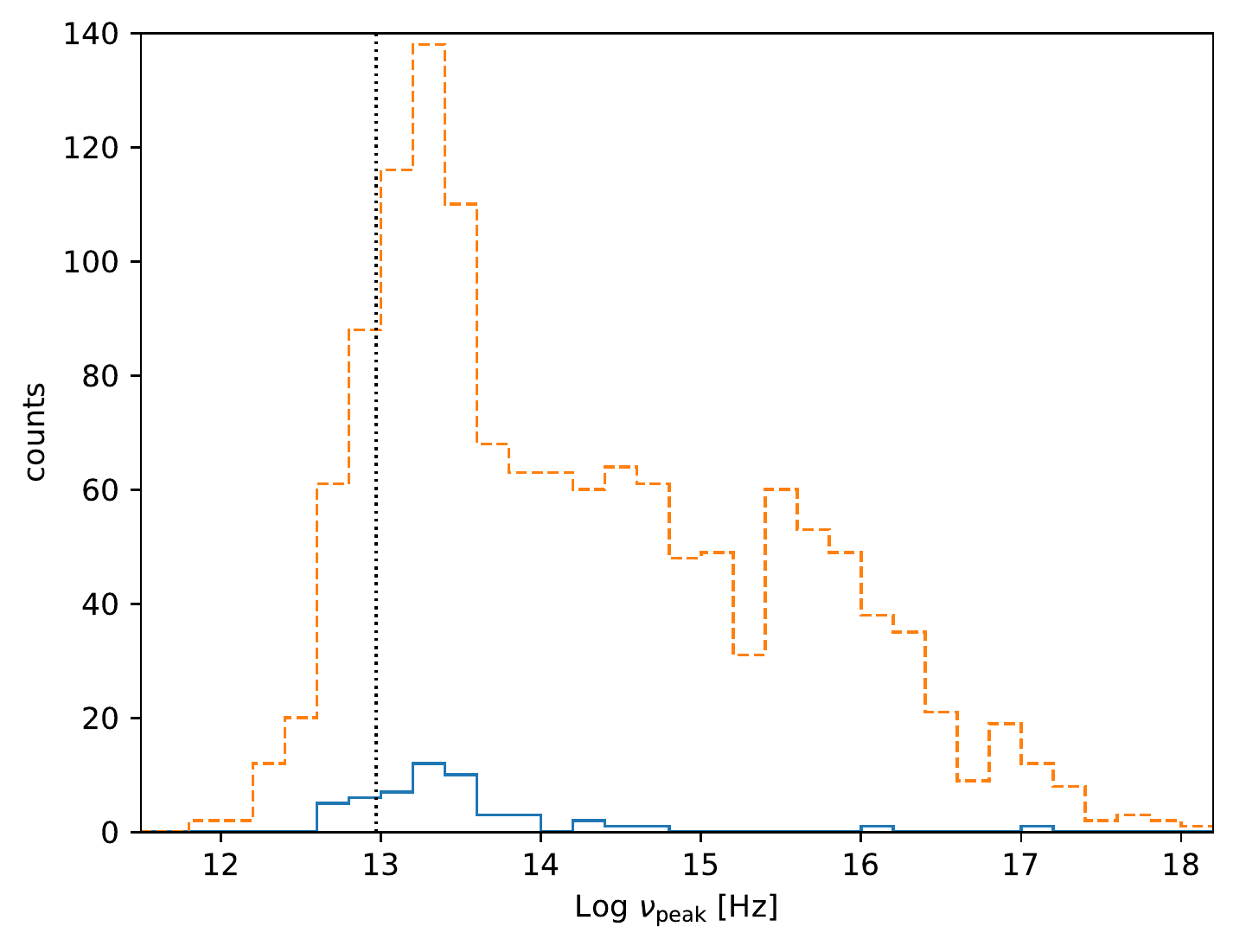}
      \caption{Histogram of the rest-frame synchrotron peak frequency for 3LAC sources (orange dashed line) and for the subset of 3LAC sources belonging to the S4 sample (blue solid line). The vertical dotted line indicates the peak frequency of \source, $\log(\nu_\mathrm{peak}/\mathrm{Hz})=12.97$.
              }
         \label{fig.sed_histo}
   \end{figure}

The sudden brightening and gamma-ray detection of \source\ is certainly in agreement with the presence of long-term high-energy variations in blazars, although a detailed gamma-ray analysis would be necessary to constrain its amplitude, which goes beyond the scope of the present paper.   On the other hand, we can discuss its SED properties and the radio variability (taken as a proxy of the Doppler factor) in comparison with the rest of the S4 blazars\footnote{Hereafter, we will only consider the 121 S4 sources appearing also in the 5th edition of the Roma-BZCat catalogue of known blazars \citep{Massaro2015}.}. 

   \begin{figure*}
   \centering
   \includegraphics[width=\hsize]{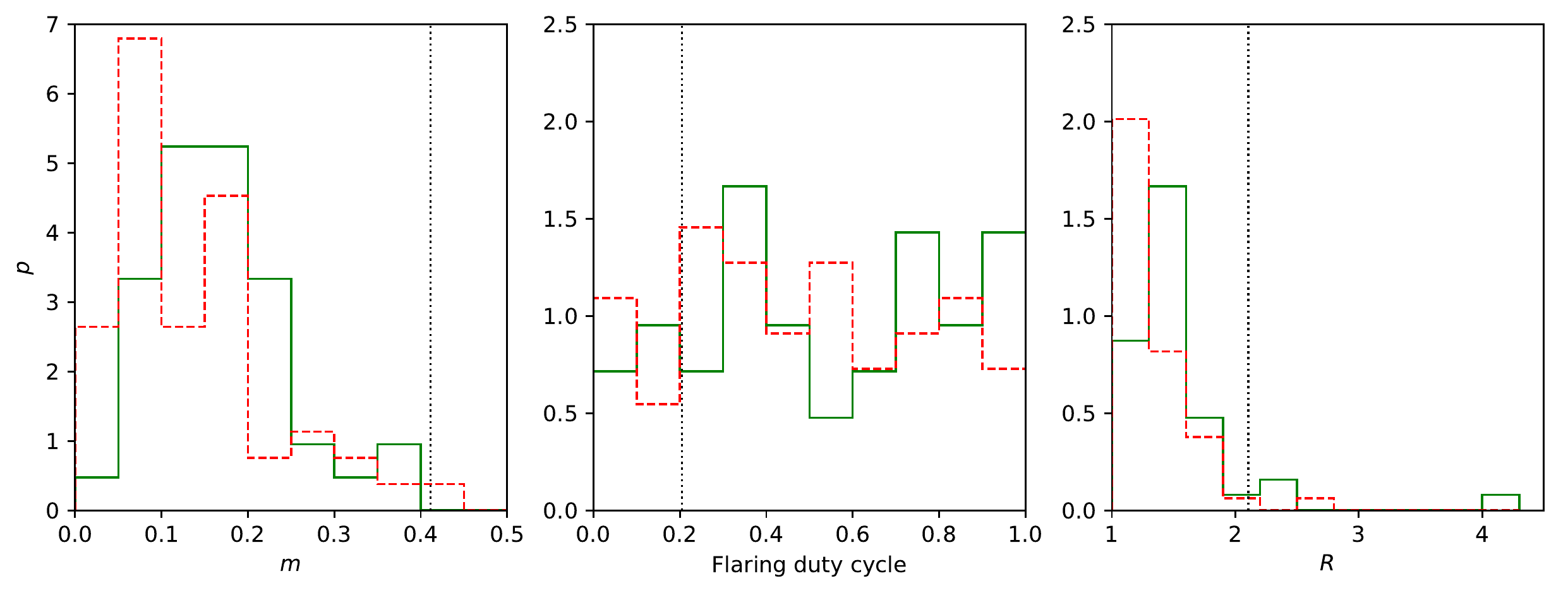}
      \caption{Probability density distribution of modulation index ($m$, left panel), flaring duty cycle (middle panel), and flaring ratio ($R$, right panel) for S4 blazars observed by the OVRO 40m telescope monitoring program, divided according to gamma-ray detection (green solid line, 42 sources) or not (red dashed line, 55 sources). The vertical dotted line indicates the value of each parameter for \source.
              }
         \label{fig.var_histo}
   \end{figure*}
   
First, we report in Fig.~\ref{fig.sed} the SED of \source\ based on data from the NASA/IPAC Extragalactic Database (NED) and from the present campaign.  A polynomial fit allows us to constrain the peak frequency of the synchrotron component as $\log(\nu_\mathrm{peak}/\mathrm{Hz})=12.97$ in the rest frame.  This value classifies the source as a low-synchrotron peaked (LSP) blazar. In Fig.~\ref{fig.sed_histo}, we further show the histogram of the peak frequency for the S4 sources detected in the 3LAC.  It is clear that, despite the significant detection of \source\ during its flare, its synchrotron peak frequency is rather low, with 87\% sources having a larger $\nu_\mathrm{peak}$ value.

Finally, we consider radio variability as a proxy of Doppler beaming, as argued by \citet{Lister2015}.  \citet{Liodakis2017} have modelled the flux density distribution of OVRO monitored blazars as a series of 'off'- and 'on'-states. By following the same approach, we considered the modulation index $m$, the flaring duty cycle, and the flaring ratio $R$ of \source\ and the other S4 sources present in the OVRO monitoring.  The corresponding distributions are plotted in Fig.~\ref{fig.var_histo}, according to their gamma-ray detection in 3LAC and indicating the values for \source.  Interestingly, with $m=0.41$ and $R=2.1$, \source\ turns out to have modulation index and flaring ratio among the highest in the sample. This suggests a relatively large Doppler factor $\delta$, probably larger than that derived from the kinematics analysis discussed in Sect.~\ref{sect.identification}.  Since the VLBI data were taken over 20 years ago, it is possible that a change in the jet bulk velocity or the viewing angle has occurred, leading to an increase in $\delta$.   It is also possible that the radio luminosity at low frequency has been significantly overestimated because of contamination from the beamed core. As recently shown by \citet{D'Antonio2019}, cores are still significantly contributing to the total radio emission at frequencies as low as 70 MHz, which could result in a systematic underestimation of the amount of beaming when using the \citet{Giovannini1988} relation. The dot-dashed line in Fig.~\ref{fig.theta_beta} indicates a possible solution for a total radio luminosity reduced by a factor of $3\times$, leading to a Doppler value $\delta=8.7$.

We can then finally conclude that \source\ fits in a picture where gamma-ray emission from sources with low synchrotron peak frequency are less likely to show bright gamma-ray emission, but large variability, likely caused by significant Doppler beaming, can lead to periods of enhanced activity during which the source becomes detectable.

\subsection{Perspectives on single dish observations}

Single dish observations, lacking detailed spatial resolution, can still be a valuable tool to investigate the physics of relativistic jets.   This is usually done exploiting long and dense monitoring campaigns for large samples of sources, preferably at multi-frequency, accompanied by dedicated statistical analysis, such as those carried out at OVRO \citep{Richards2011,Max-Moerbeck2014}, Effelsberg \citep{Fuhrmann2014,Angelakis2019}, Metsah\"ovi \citep{Hovatta2008}, to name a few.
The Medicina and Noto radio telescopes are also involved in a similar program, involving over 30 blazars being monitored regularly between 5 and 43 GHz for the past 1.5 decades \citep{Bach2007}. The development of CAP is expected to provide a significant boost to the scientific return of the observational efforts.  A preliminary overview is given in \citet{Righini2019}.

Smaller-scope projects such as the ones presented in this work are however useful in that they can provide a characterisation of sources over short time scales. Other examples benefitting from single dish data analysed with CAP include the recent identification of PKS\,1153--1105 as a new gamma-ray blazar \citep{Giroletti2018} or the monitoring of transient events from galactic sources undergoing rapid evolution such as Cyg X-3 \citep{Egron2017}.

Along with the development of the radio telescope, which is being provided with additional devices such as new digital back-ends for fast spectral acquisitions, our data reduction software is planned to evolve, too. The availability of OTF spectral data will allow the continuum measurements to be performed on frequency-integrated spectra after proper RFI-excision operations, thus our code will have to provide such features or, at least, be able to comply with the requirements of other dedicated software. This first version of CAP still needs intervention by part of the user, to flag the data and to handle those cases in which acquisitions are only partially usable - e.g when cross-scans are incomplete. We aim to implement new features in order to automatise such procedures as much as possible.

\section{Conclusions}
\label{sect.conclusions}

Following a gamma-ray flare, we observed \source\ in the radio for almost a year.  The source has shown to be in elevated state with respect to historical values since the beginning of our campaign. It reached a peak luminosity of $L=(1.7\pm0.3)\times10^{27}\ \mathrm{W\,Hz}^{-1}$ at 24 GHz and then slowly decreased almost to its historical state. The spectral index has remained consistent with flat throughout the observations, although it has shown an evolution from slightly inverted values to slightly positive.

The observations reported here have thus been instrumental in providing temporal and spectral information in the radio domain for the gamma-ray source reported by \citet{Ciprini2017}, pointing to an identification of the radio and gamma-ray sources.  In comparison with other gamma-ray blazars, \source\ has relatively low synchrotron peak frequency,   which makes it hard to detect at GeV energies in low state, and large Doppler factor, which by contrast can result in periods of enhanced activity leading the detection in gamma rays.

Beside the implications for the physics of the source itself, the current observations have also demonstrated the efficiency and the reliability of the new calibration and data analysis of data from the INAF single dish radio telescopes. Work is ongoing to reduce and interpret a vast amount of data collected over the last years; observations and software development are also being carried out in order to improve the scientific return of the facilities.

\section*{Acknowledgements}

We thank an anonymous referee for very constructive suggestions which significantly improved the manuscript.
We thank P.~Cassaro and P.~R.~Platania for carrying out the observations at the Noto radiotelescope. Based on observations with the Medicina and Noto telescopes operated by INAF - Istituto di Radioastronomia. This research has made use of the NASA/IPAC Extragalactic Database (NED) which is operated by the Jet Propulsion Laboratory, California Institute of Technology. This research has made use of data from the OVRO 40-m monitoring program (Richards, J. L. et al. 2011, ApJS, 194, 29) which is supported in part by NASA grants NNX08AW31G, NNX11A043G, and NNX14AQ89G and NSF grants AST-0808050 and AST-1109911.

\appendix
\section{On the accuracy of flux density measurements with CAP}
\label{sect.appendix}

   \begin{figure*}
   \centering
   \includegraphics[width=0.67\textwidth]{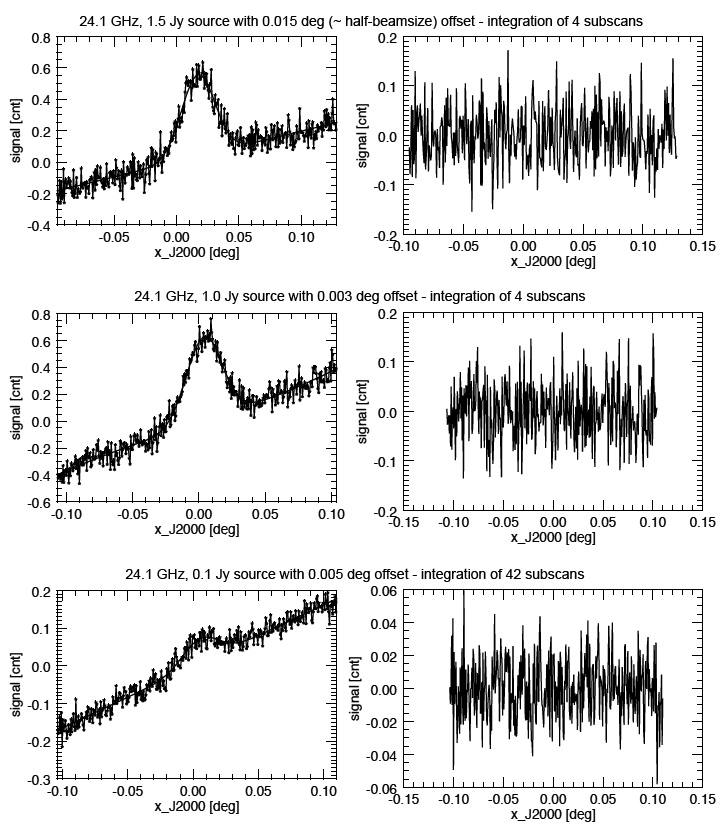}
      \caption{Examples of integrations of simulated subscans. Synthetic sources of different flux densities were generated and, after having applied the effects due to pointing errors and atmospheric opacity to their amplitudes, added to a white noise corresponding to the typical observing setups and conditions. The left panels show counts (as connected dots) overlaid with a Gaussian plus linear fit; the right panels show the corresponding residuals.}
         \label{fig.synthplots}
   \end{figure*}
   
The final flux density measurements reported by CAP are the result of several steps, each one contributing to the total error budget.  In this Appendix, we describe how we estimate the final uncertainty associated to flux densities (\ref{A1}), and then present some tests performed against synthetic data (\ref{A2}) and using real data acquired on flux density calibrators (employing them as if they were unknown target sources, \ref{A3}).

\subsection{Estimate of uncertainty}\label{A1}
For each calibrator scan, the Gaussian plus sky baseline (either linear or polynomial) fit to the gain- (and opacity-, when applicable) corrected counts returns an uncertainty $$\sigma_{\mathrm{fit},\phi;i,j,k}, \ i=0,1, \ j=0,1, \ k=0,\dots,n$$
with $i$ representing the two polarisation states ($i=0$ for LCP, $i=1$ for RCP), $j$ the two scan directions ($j=0$ for a scan in right ascension $\alpha$, $j=1$ for one in declination $\delta$), and $k$ the calibration scan (with $n$ being the total number of scans); the subscript $\phi$ indicates that these quantities are still affected by a possible pointing offset.   This uncertainty is then propagated to the counts-to-Jy calibration factor $q_{\phi;i,j,k}=S/A_{\mathrm{fit},\phi;i,j,k}$ (where $S$ is the calibrator flux density, known a-priori, and $A_{\mathrm{fit},\phi;i,j,k}$ is the amplitude of the fit in counts), given by
$$\sigma_{q,\phi;i,j,k}=q_{\phi;i,j,k}\times\sigma_{\mathrm{fit},\phi;i,j,k}/A_{\mathrm{fit},\phi;i,j,k}$$

The calibration factors and associated uncertainties determined for each scan along the two directions are then
\begin{itemize}
    \item[--] rescaled according to the measured $\Delta\phi_{i,1-j,k}$; i.e.\ the offset determined in the orthogonal (i.e.\ $1-j$) direction for the same polarisation $i$ and scan $k$
$$\sigma_{q;i,j,k}=\sigma_{q,\phi;i,j,k} \times \exp{(-1.66\Delta\phi_{i,1-j,k}/\theta)^2}$$
($\theta$ being the angular size of the antenna primary beam)
    \item[--] combined with a weighted mean, i.e.\ for each polarisation $i$ and scan $k$ the uncertainty is
$$\sigma_{\mathrm{q};i,k}=\frac{\sqrt{\sum_{j=0,1}w^2_{i,j,k}\sigma^2_{\mathrm{fit};i,j,k}}}{\sum_{j=0,1}w_{i,j,k}}$$
where the weights $w_{i,j,k}$ are the inverse of the relative error at each $(i,j,k)$ combination.
\end{itemize}

As described in Sect.~\ref{sect.observations}, if there is more than one calibrator scan (i.e.\ if $k>1$), the calibration factors are linearly interpolated taking account of their uncertainties $\sigma_{\mathrm{fit},i,k}$ returning a time dependent calibration factor with associated uncertainty $\sigma_{q-i}(t)$. 

For target sources, we also start from the Gaussian plus sky baseline (either linear or polynomial) fit to the gain- (and opacity-, when applicable) corrected count uncertainty
$$\sigma'_{\mathrm{fit},\phi;i,j,k}, \ i=0,1, \ j=0,1, \ k=0,\dots,n$$
which in this case is immediately corrected for the pointing offsets
$$\sigma'_{\mathrm{fit};i,j,k}=\sigma'_{\mathrm{fit},\phi;i,j,k} \times \exp{(-1.66\Delta\phi_{i,1-j,k}/\theta)^2}$$
and converted to flux densities by taking into account the calibration factor at the time $t'$ of the target scan, resulting in an uncertainty for each receiver polarisation $i$, scan direction $j$, and target scans $k$
$$\sigma'_{i,j,k}=S'\times \sqrt{\left(\frac{\sigma'_{\mathrm{fit};i,j,k}}{A'_{\mathrm{fit};i,j,k}}\right)^2+\left(\frac{\sigma_{q-i}(t')}{q_i(t')}\right)^2}$$

Through subsequent weighted means, the different values of $i, j, k$  are combined in the final flux density $S'$ and associated uncertainty $\sigma'$. 

\subsection{Test against synthetic data}\label{A2}

\begin{figure*}
   \centering
   \includegraphics[width=\columnwidth]{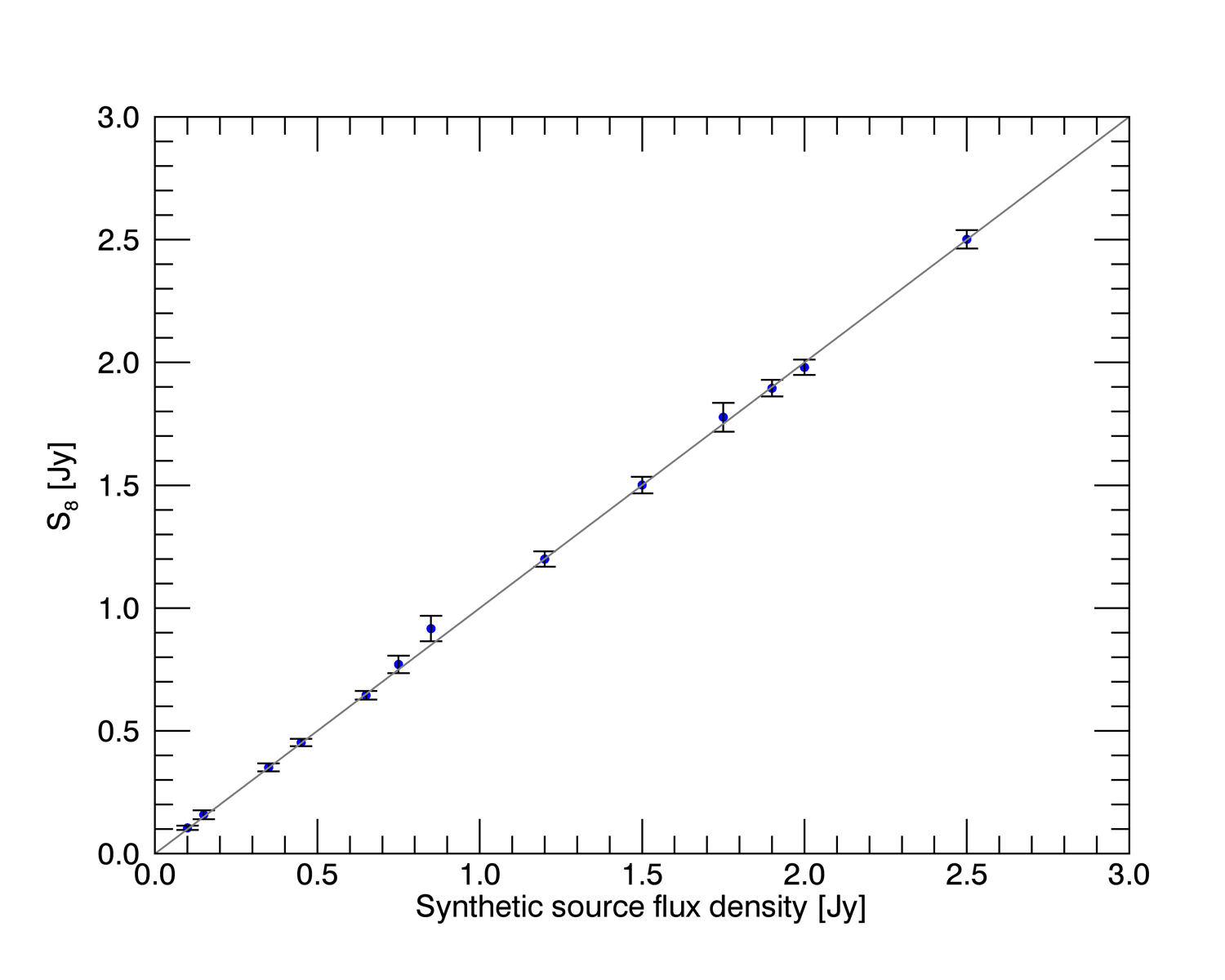}
   \includegraphics[width=\columnwidth]{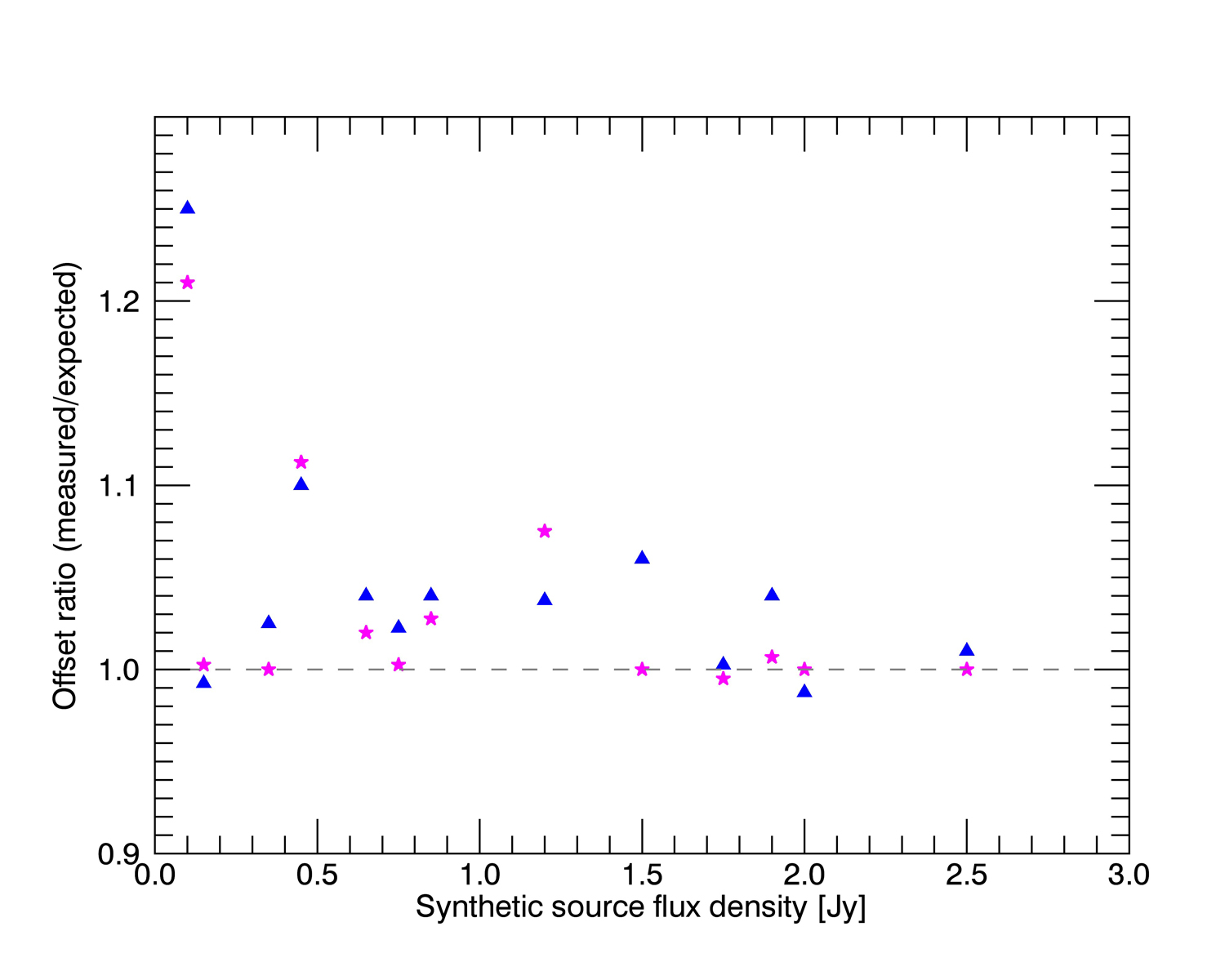}
      \caption{Comparison between measured and expected quantities for 8-GHz simulated data. Left: measured vs simulated flux density; right: pointing offset ratio (blue triangles = RA offset, magenta stars = Dec offset).}
         \label{fig.xratio}
   \end{figure*}

\begin{figure*}
   \centering
   \includegraphics[width=\columnwidth]{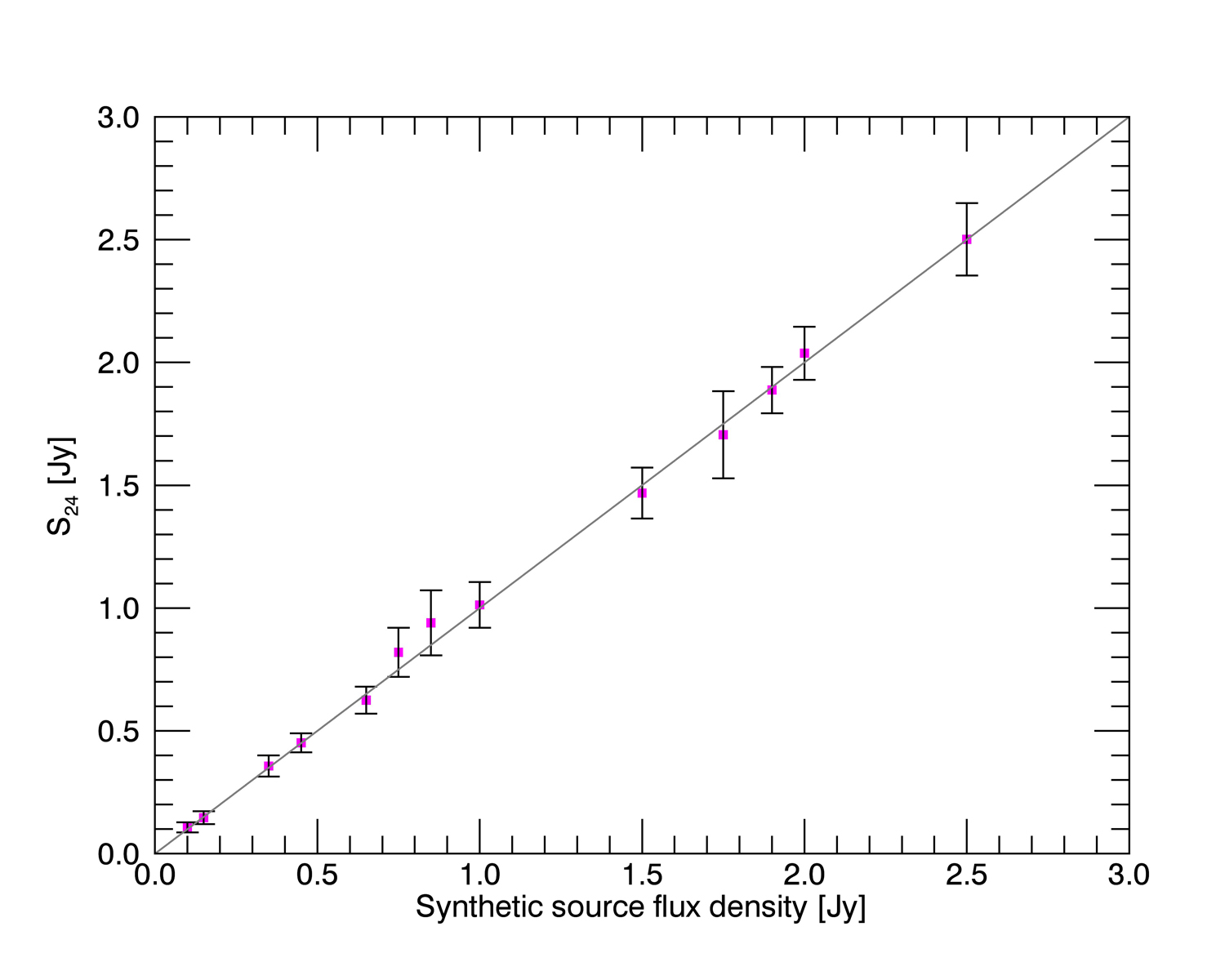}
   \includegraphics[width=\columnwidth]{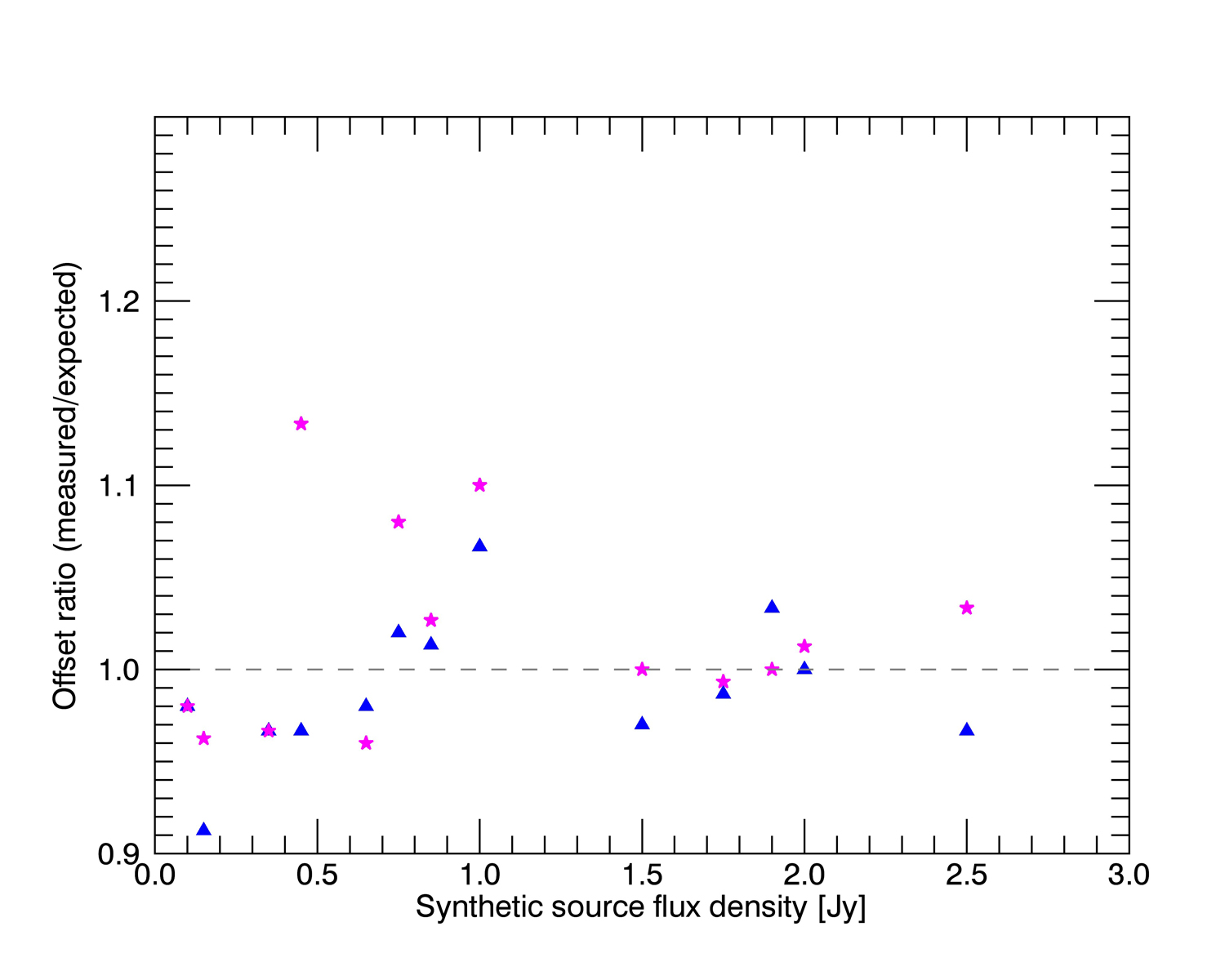}
      \caption{Comparison between measured and expected quantities for 24-GHz simulated data. Left: measured vs simulated flux density; right: pointing offset ratio (blue triangles = RA offset, magenta stars = Dec offset).}
   \label{fig.kratio}
   \end{figure*}

Synthetic cross-scans were built in order to reproduce a clean version - i.e. RFI-free - of the typical acquisitions performed both at 8 and 24~GHz. Real-life parameters - in terms of antenna gain curve, beamsize and Tsys - were employed together with the cross-scan instrumental and geometrical setup used within our project. For 24-GHz data, the presence of different atmospheric conditions was also simulated. 
In order to test the software capability to correctly estimate the pointing offset, too, we generated these fake sources rescaling their apparent amplitude according to a variety of pointing offset values (see examples in Fig.~\ref{fig.synthplots}). We then used part of the simulated sources as flux density calibrators, while the others played the role of unknown target sources. Results are shown in Figs.~\ref{fig.xratio}~and~\ref{fig.kratio}. 24~GHz data, having higher rms-noise due to the greater Tsys (100~K w.r.t. 40~K) and being complicated by the injection of atmospheric opacity, produce measurements with larger uncertainties. Pointing offsets, as expected, turn out to be more precisely measured when the signal-to-noise ratio increases.

      \begin{figure*}
   \centering
   \includegraphics[width=\columnwidth,trim=90 220 90 220,clip]{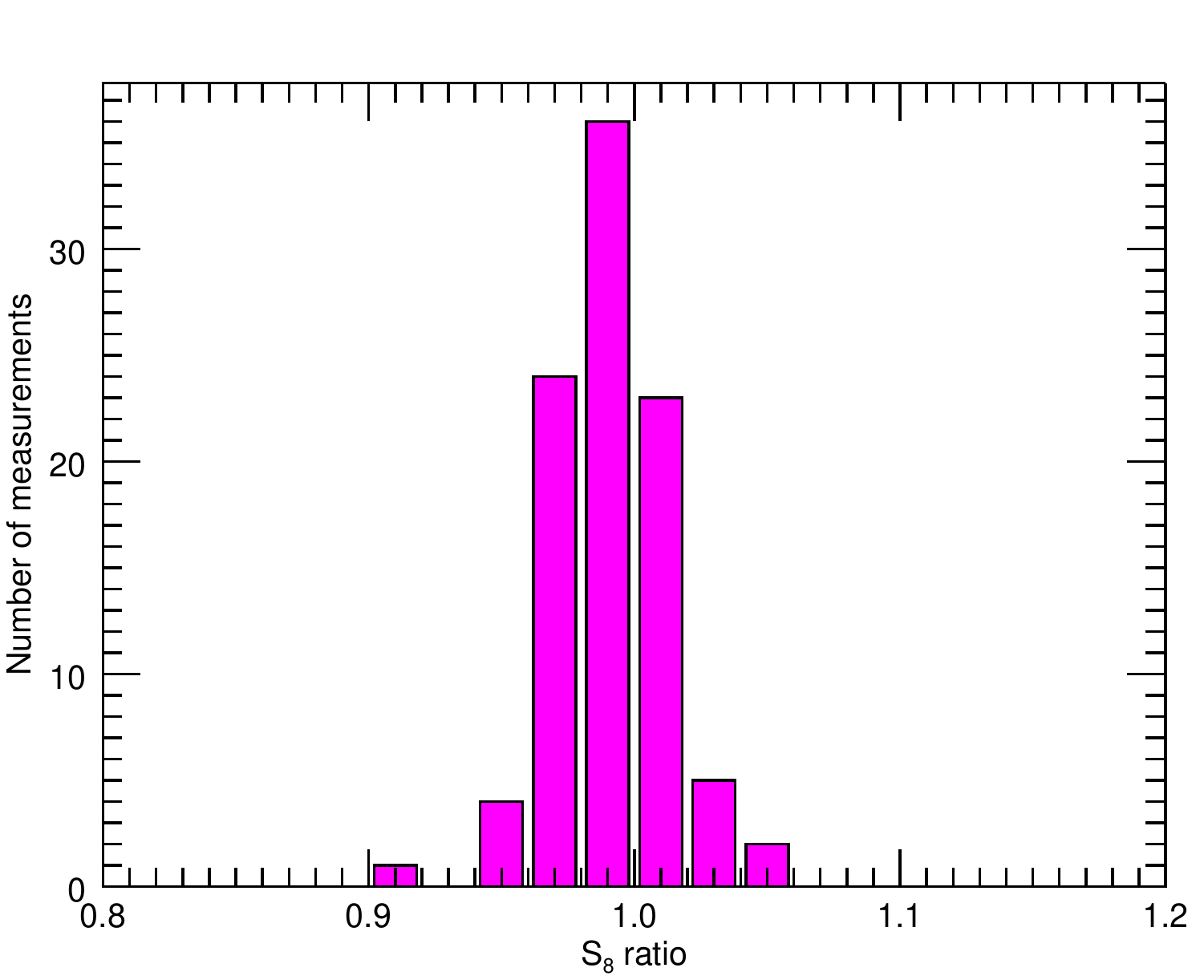}   \includegraphics[width=\columnwidth,trim=90 220 90 220,clip]{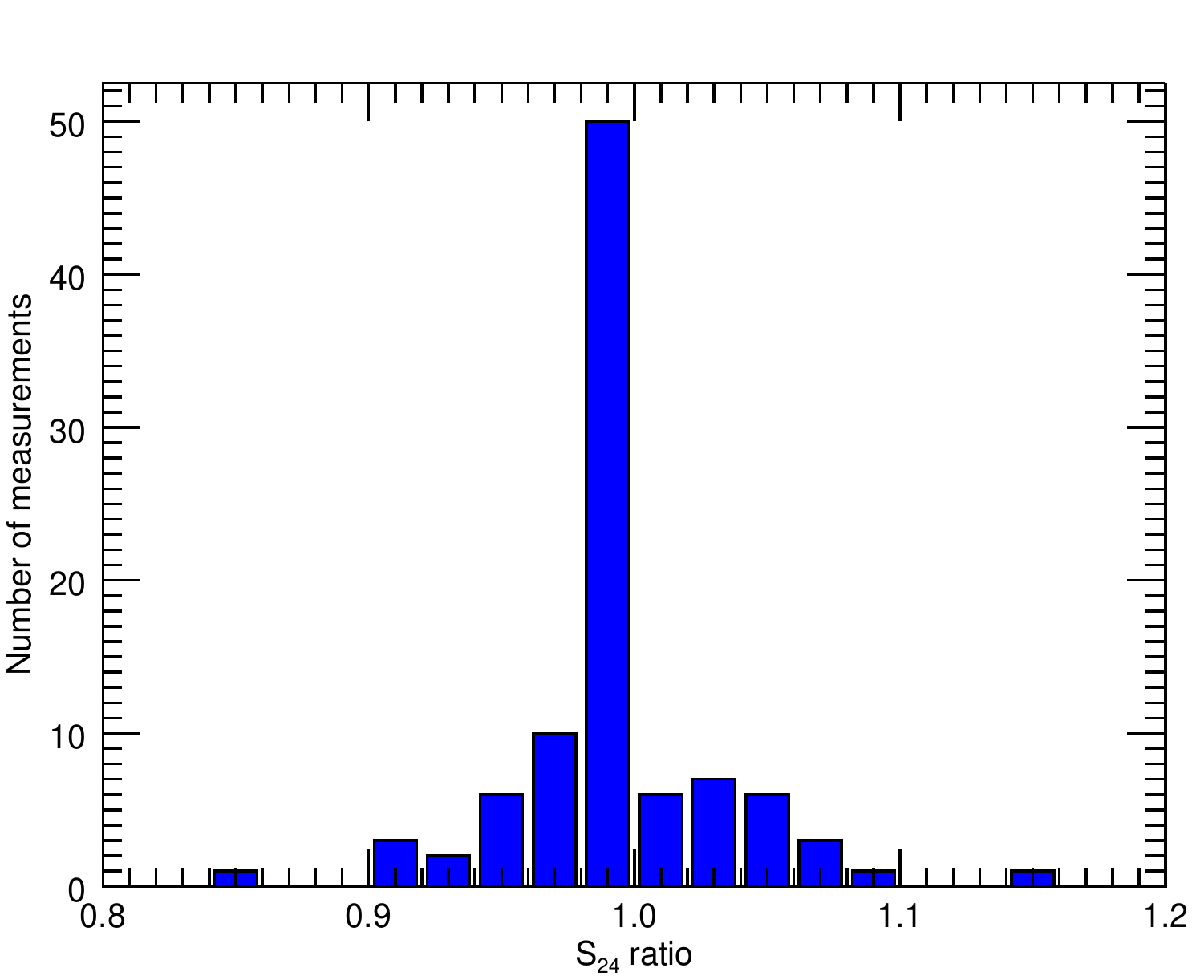}
      \caption{Distribution of the flux density ratio (measured/expected) for 8- and 24-GHz real acquisitions on flux density calibrators treated as target sources.}
         \label{fig.calgets}
   \end{figure*}

\subsection{Tests based on real data}\label{A3}
    
The second test consisted in gathering real data, acquired on flux density calibrators, from several observing sessions, and reduce them as if such sources were also normal targets. Matching the procedures we usually follow in reducing our data, flux density calibration was performed applying counts-to-Jy factors averaged over 1 hour for 24-GHz simulated scan, and over 24 hours for the 8-GHz ones. In order to perform a more realistic test, as the sources both acted as calibrators and targets, for each source we split the acquired subscans in two independent tranches, using the first for the extraction of the calibration factors and the second for the flux density measurements. This implied that the on-source integration was half of the usually-employed one, reducing the signal-to-noise ratio by $1/\sqrt{2}$. As these calibration sources are bright and observations were slightly redundant, this did not cause major difficulties in the detection. Histograms in Figure \ref{fig.calgets} show how performing the measurements were. The better results at 24~GHz are likely due to the acquisitions being cleaner (i.e. with negligible RFI contributions) than 8~GHz ones. Moreover, our high-frequency observations always include, prior to acquisitions, a fine-pointing procedure that practically zeroes the pointing offsets - and the inaccuracies that their compensation implies within our reduction pipeline.







\bsp	
\label{lastpage}
\end{document}